\documentclass[doublecol]{epl2}

\usepackage{graphicx}
\usepackage{dcolumn}
\usepackage{bm}

\title{Superconductivity in LaFeAs$_{1-x}$P$_{x}$O: effect of chemical pressures and bond covalency}
\shorttitle{Superconductivity in LaFeAs$_{1-x}$P$_{x}$O}

\author{Cao Wang,\inst{1} Shuai Jiang,\inst{1} Qian Tao,\inst{1} Zhi Ren,\inst{1} Yuke Li,\inst{1} Linjun Li,\inst{1} Chunmu Feng,\inst{2}  Jianhui Dai,\inst{1} Guanghan Cao\inst{1}\footnote{Electronic address: ghcao@zju.edu.cn} and Zhu-an Xu\inst{1}\footnote{Electronic address: zhuan@zju.edu.cn}}
\shortauthor{C. Wang \etal}

 \institute{
  \inst{1} Department of Physics, Zhejiang University - Hangzhou 310027, China\\
  \inst{2} Test and Analysis Center, Zhejiang University - Hangzhou 310027, China
}


\abstract{We report the realization of superconductivity by an
isovalent doping with phosphorus in LaFeAsO. X-ray diffraction shows
that, with the partial substitution of P for As, the Fe$_2$As$_2$
layers are squeezed while the La$_2$O$_2$ layers are stretched along
the $c$-axis. Electrical resistance and magnetization measurements
show emergence of bulk superconductivity at $\sim$10 K for the
optimally-doped LaFeAs$_{1-x}$P$_{x}$O ($x=0.25\sim0.3$). The upper
critical fields at zero temperature is estimated to be 27 T, much
higher than that of the LaFePO superconductor. The occurrence of
superconductivity is discussed in terms of chemical pressures and
bond covalency.}

\begin{document}

\maketitle

\section{Introduction}Superconductivity can be induced by carrier doping in an insulator,
semiconductor, and even metal. Representative examples are shown in
hole-doped La$_2$CuO$_4$\cite{Bednorz&Muller}, electron-doped
BaBiO$_3$\cite{Cava1988}, and electron-doped
TiSe$_{2}$\cite{Cava2006}. Recently, superconductivity at 26 K was
discovered in LaFeAsO by either electron doping with
florine\cite{Hosono} or hole doping with strontium\cite{Wen}.
Subsequent replacements of La with other rare earth elements raised
the critical temperatures ($T_c$) over the McMillan limit (39
K).\cite{Chen-Sm,Chen-Ce,Ren-Sm} By electron doping with thorium in
GdFeAsO, $T_c$ has reached 56 K.\cite{Wang} The discovery of high
superconducting transition temperatures in these Fe-based compounds
has generated great interest in the scientific
community.\cite{Johannes}

As a prototype parent compound of the new class of high-temperature
superconductors, LaFeAsO crystallizes in ZrCuSiAs-type
structure,\cite{Quebe} which consists of insulating
[La$_2$O$_2$]$^{2+}$ layers and conducting [Fe$_2$As$_2$]$^{2-}$
layers. In addition to the carrier doping in [La$_2$O$_2$]$^{2+}$
layers, partial substitution of Fe with Co\cite{Oak-Co,Wang-Co} and
Ni\cite{Ni} also leads to superconductivity. Although the valence of
the doped Co and Ni seems to remain 2+, electron carriers were
believed to be induced owing to the itinerant character of the 3$d$
electrons.\cite{Wang-Co} That is to say, the Fe-site substitution by
Co/Ni still belongs to the scenario of carrier doping.

Apart from chemical doping, superconductivity was also observed via
applying hydrostatic pressure in the parent compounds such as
$A$Fe$_2$As$_2$ ($A$=Ca, Sr, Ba and
Eu)\cite{P-Ca122,P-SrBa122,P-Eu122} and LaFeAsO\cite{P-La1111}. As
"chemical pressures" may be produced by an isovalent substitution
with smaller ions, we have tried the substitution of As by P in
EuFe$_{2}$As$_{2}$.\cite{Eu122P} As a result, superconductivity
appears below 26 K. Nevertheless, the superconductivity is then
influenced by the subsequent ferromagnetic ordering of Eu$^{2+}$
moments, and diamagnetic Meissner effect cannot be observed.

LaFeAsO is a prototype parent compound of ferroarsenide
superconductors, showing spin-density-wave (SDW) antiferromagnetic
ground state.\cite{DaiPC} In contrast, the other end member
LaFeAs$_{1-x}$P$_{x}$O ($x$=1) is a superconductor of $\sim$4 K,
showing non-magnetic behavior in the normal state.\cite{LaFePO}
According to a recent theory,\cite{JDai} partial substitution of P
for As in the ferroarsenides may induce a quantum criticality, which
could induce superconductivity. Therefore, the effect of P doping in
LaFeAsO is of great interest. In this Letter, we demonstrate bulk
superconductivity in LaFeAs$_{1-x}$P$_{x}$O at $\sim$10 K with the
evidences of both zero resistance and Meissner effect. This result
establishes a stronger evidence that "chemical pressures" and/or
bond covalency may stabilize superconductivity in the ferroarsenide
system.

\section{Experimental}Polycrystalline samples of LaFeAs$_{1-x}$P$_{x}$O were synthesized
by solid state reaction in vacuum using powders of LaAs,
La$_{2}$O$_{3}$, FeAs, Fe$_{2}$As, FeP and Fe$_{2}$P. Similar to our
previous report,\cite{Wang-Co} LaAs, FeAs, Fe$_{2}$As, FeP and
Fe$_{2}$P were presynthesized, respectively. La$_{2}$O$_{3}$ was
dried by firing in air at 1173 K for 24 hours prior to using. All
the starting materials are with high purity ($\geq$ 99.9\%). The
powders of these intermediate materials were weighed according to
the stoichiometric ratios of LaFeAs$_{1-x}$P$_{x}$O ($x$=0, 0.1,
0.2, 0.25, 0.3, 0.35, 0.4, 0.5 and 0.6), thoroughly mixed in an
agate mortar, and pressed into pellets under a pressure of 2000
kg/cm$^{2}$, operating in a glove box filled with high-purity argon.
The pellets were sealed in evacuated quartz tubes, then heated
uniformly at 1373 K for 40 hours, and finally furnace-cooled to room
temperature.

Powder x-ray diffraction (XRD) was performed at room temperature
using a D/Max-rA diffractometer with Cu-K$_{\alpha}$ radiation and a
graphite monochromator. The detailed structural parameters were
obtained by Rietveld refinements, using the step-scan XRD data with
$10^{\circ}\leq 2\theta \leq 120^{\circ}$.

The electrical resistivity was measured using a standard
four-terminal method. The measurements of magnetoresistance and Hall
coefficient were carried out on a Quantum Design physical property
measurement system (PPMS-9). The measurements of dc magnetic
properties were performed on a Quantum Design Magnetic Property
Measurement System (MPMS-5). Both the zero-field-cooling (ZFC) and
field-cooling (FC) protocols were employed under the field of 10 Oe.

\section{Results and discussion}
\begin{figure}
\onefigure[width=8cm]{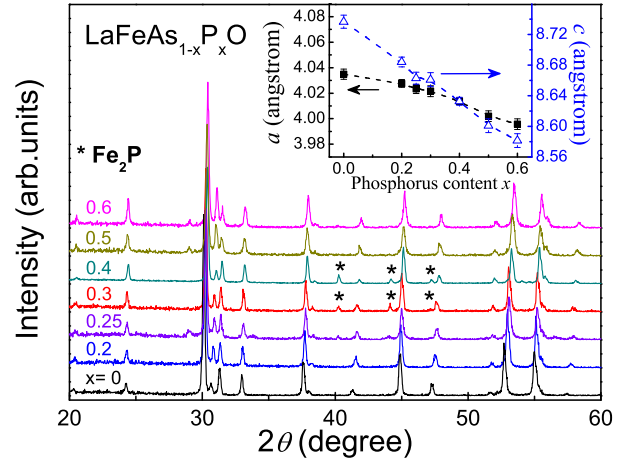}\caption{X-ray powder diffraction
patterns at room temperature for the LaFeAs$_{1-x}$P$_{x}$O samples.
Small amount of Fe$_{2}$P impurity is marked by asterisks. The inset
plots the lattice parameters as functions of nominal phosphorus
content.} \label{fig.1}
\end{figure}

Figure ~\ref{fig.1} shows the XRD patterns for
LaFeAs$_{1-x}$P$_{x}$O samples. The sample of $x$=0 shows single
phase of LaFeAsO. With the P doping over 20\%, small amount of
Fe$_2$P impurity appears. When the doping level exceeds 50\%,
however, the impurity phase tends to disappear. The inset of Fig.
~\ref{fig.1} plots the calculated lattice parameters as functions of
nominal P content. Both \emph{a}-axis and \emph{c}-axis decrease
with increasing $x$. Compared with the undoped LaFeAsO,
\emph{a}-axis decreases by 0.34\% while \emph{c}-axis shrinks by
0.87\% for LaFeAs$_{0.7}$P$_{0.3}$O. Thus the isovalent substitution
of As with P indeed generates chemical pressure to the system. We
note that the shrinkage in the basal planes is similar to that in
EuFe$_{2}$(As$_{1-x}$P$_{x}$)$_2$, but the compression along
$c$-axis is not as large as that in
EuFe$_{2}$(As$_{1-x}$P$_{x}$)$_2$.\cite{Eu122P}

\begin{figure}
\includegraphics[width=8cm]{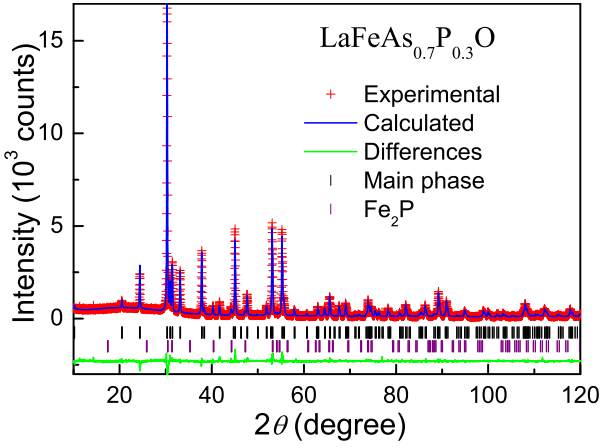}
\caption {An example of Rietveld refinement profile for
LaFeAs$_{0.7}$P$_{0.3}$O. The Fe$_2$P impurity was included in the
refinement.} \label{fig.2}
\end{figure}

The crystallographic parameters were obtained by the Rietveld
refinement based on the ZrCuSiAs-type structure. An example of the
refinement is seen in Fig.~\ref{fig.2}. The reliability factor
$R_{wp}$ is 8.9\% and the goodness of fit is 1.7, indicating fairly
good refinement for the crystallographic parameters.
Table~\ref{tab.1} compares the structural data of the undoped and
P-doped (by 30 at.\%) samples. It is clear that As/P atoms are
closer to the Fe planes for the P-doped compound, resulting in the
flattening of the Fe$_{2}$As$_{2}$ layers. Interestingly, La atoms
move toward the Fe$_{2}$As$_{2}$ layers, leading to the
\emph{increase} of the La$_{2}$O$_{2}$ layers. Thus the chemical
pressure induced by the P doping actually causes compression in
Fe$_{2}$As$_{2}$ layers, but stretching in La$_{2}$O$_{2}$
thickness, along the $c$-axis. This explains why the decrease in
$c$-axis in LaFeAs$_{1-x}$P$_{x}$O is not as so much as that in
EuFe$_{2}$(As$_{1-x}$P$_{x}$)$_2$. Besides, with the flattening of
the Fe$_{2}$As$_{2}$ layers, the bond angle of As-Fe-As increases
obviously. The large As-Fe-As angle may account for the relatively
low $T_c$ in LaFeAs$_{1-x}$P$_{x}$O system (to be shown below),
according to the empirical structural rule for $T_c$ variations in
ferroarsenides.\cite{Structure-Tc}.

\begin{table}
\caption{Crystallographic data of LaFeAs$_{1-x}$P$_{x}$O ($x$=0 and
0.3) at room temperature. The space group is P$4/nmm$. The atomic
coordinates are as follows: La (0.25,0.25,$z$); Fe (0.75,0.25,0.5);
As/P (0.25,0.25,$z$); O (0.75,0.25,0).} \label{tab.1}
\begin{center}
\begin{tabular}{lcr}
\hline
Compounds&LaFeAsO&LaFeAs$_{0.7}$P$_{0.3}$O\\
\hline
$a$ ({\AA}) & 4.0357(3) &4.0219(1)\\
$c$ ({\AA}) & 8.7378(6) &8.6616(3)\\
$V$ ({\AA}$^3$) & 142.31(2) & 140.10(1)\\
$z$ of La & 0.1411(2)& 0.1435(1)\\
$z$ of As & 0.6513(3) & 0.6475(3)\\
La$_{2}$O$_{2}$ thickness ({\AA}) & 2.466(2) &2.486(1)\\
Fe$_{2}$As$_{2}$ thickness ({\AA}) & 2.644(2) &2.555(1)\\
Fe-Fe spacing ({\AA}) & 2.8536(3) &2.8439(1)\\
As-Fe-As angle ($^{\circ}$) & 113.5(1) &115.1(1)\\
\hline
\end{tabular}
\end{center}
\end{table}

\begin{figure}
\onefigure[width=8cm]{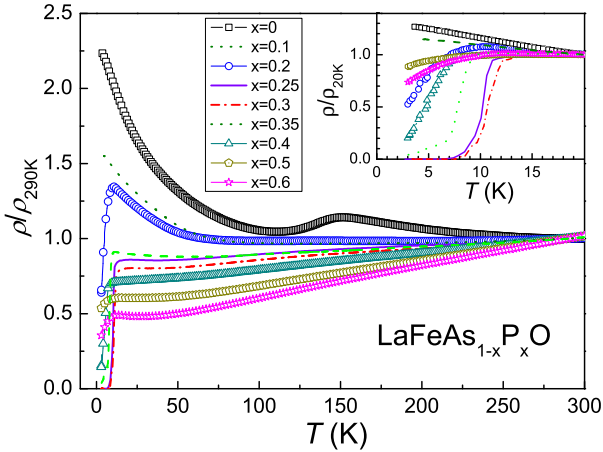} \caption{Temperature dependence of
resistivity for LaFeAs$_{1-x}$P$_{x}$O samples. The inset shows an
expanded plot. The data are normalized for comparison.}\label{fig.3}
\end{figure}

The temperature dependence of resistivity ($\rho$) for
LaFeAs$_{1-x}$P$_{x}$O samples is shown in Fig.~\ref{fig.3}. The
$\rho$ of the undoped LaFeAsO show anomaly at 150 K, where a
structural phase transition takes place.\cite{DaiPC} On doping 10\%
P, the anomaly is hardly to see and $\rho$ exhibits semiconducting
behavior below 100 K. For $x$=0.2, $\rho$ is found to decrease
quickly below 7 K, suggesting a superconducting transition though
zero resistance is not achieved down to 3 K. For \emph{x}=0.25 and
0.3, $\rho$ drops sharply at $\sim$ 11 K, and the midpoint
superconducting transition temperatures $T_{c}^{mid}$ are 10.3 K and
10.8 K, respectively. With further increasing \emph{x} to 0.4,
$T_{c}^{mid}$ decreases to 5 K, and the superconducting transition
becomes broadened. For $x$=0.5 and 0.6, only kinks are shown below
10 K in the $\rho(T)$ curves, suggesting that the superconducting
phase has a very small fraction.

\begin{figure}
\onefigure[width=8cm]{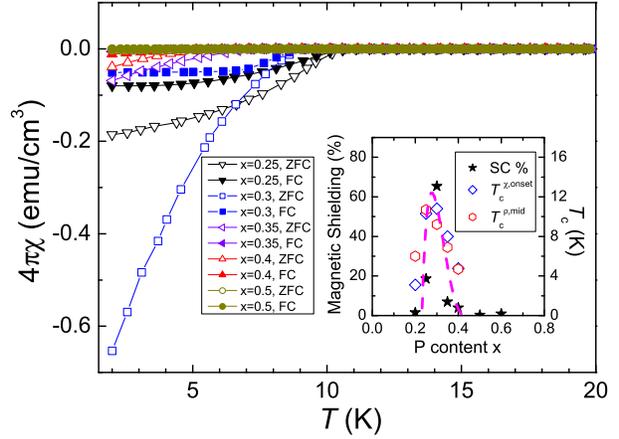} \caption{Temperature dependence of
dc magnetic susceptibility of LaFeAs$_{1-x}$P$_{x}$O (0.25$\leq
x\leq$0.6) samples. The applied field is 10 Oe. Note that the
background signals due to ferromagnetic Fe$_2$P impurity was
deducted to show the superconducting transitions clearly. The inset
shows the superconducting transition temperature and the
superconducting magnetic shielding percentage as functions of doping
level $x$.}\label{fig.4}
\end{figure}

Superconductivity in LaFeAs$_{1-x}$P$_{x}$O is confirmed by the dc
magnetic susceptibility ($\chi$) measurements, shown in
figure~\ref{fig.4}. After deducted the magnetic background signals
of the ferromagnetic Fe$_2$P impurity, diamagnetic transitions are
obvious for the superconducting samples. Strong diamagnetic signals
can be seen below 11 K for samples of $x$=0.25 and 0.3 in both ZFC
and FC data. The volume fraction of magnetic shielding (ZFC) at 2 K
achieves 65\% for $x$=0.3, indicating bulk superconductivity. For
the samples of $x$=0.2 and 0.4, however, the magnetic shielding
fraction at 2 K is less than 2\% and 5\%, respectively, suggesting
inhomogeneity of the P doping for these two samples. The samples of
$x$=0.5 and 0.6 only show trace and broad diamagnetic signals,
consistent with the kinks in above resistivity measurements. Thus
one would expect non-superconductivity for a uniform samples of
$x\leq 0.2$ and $x \geq0.4$. The superconducting phase diagram is
depicted in the inset of Fig.~\ref{fig.4}. A dome-like $T_{c}(x)$
curve is displayed.

\begin{figure}
\onefigure[width=8cm]{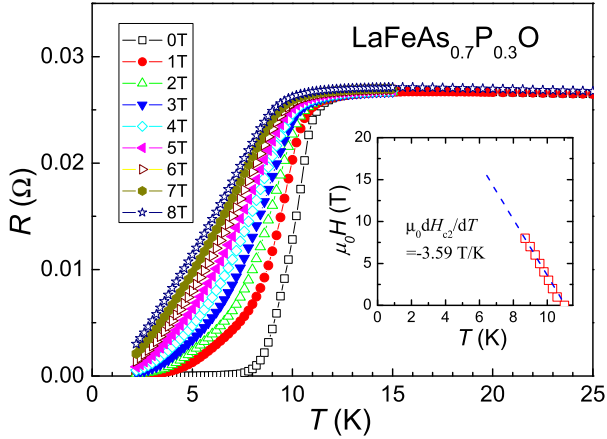} \caption{Temperature dependence of
the electrical resistance of the LaFeAs$_{0.7}$P$_{0.3}$O sample
around $T_c$ in fixed applied magnetic fields. The inset shows the
temperature dependence of the upper critical magnetic
fields.}\label{fig.5}
\end{figure}

Fig.~\ref{fig.5} shows the temperature dependence of resistivity
under magnetic fields for LaFeAs$_{0.7}$P$_{0.3}$O. As expected, the
resistive transition shifts towards lower temperature with
increasing magnetic fields. The broad transition tails under
magnetic fields are probably due to the superconducting weak links
in grain boundaries as well as the vortex motion. Thus we define
$T_{c}(H)$ as a temperature where the resistivity falls to 90\% of
the normal-state value. The initial slope $\mu_{0}$d$H_{c2}$/d$T$
near $T_{c}$ is $-$3.59 T/K, shown in the inset of Fig.~\ref{fig.5}.
The upper critical field $\mu_{0}H_{c2}$(0) is then estimated to be
$\sim$ 27 T by using the Werthamer-Helfand-Hohenberg (WHH) relation,
$H_{c2}$(0)$\approx$ 0.691$\mid$d$H_{c2}$/dT$\mid$$T_{c}$.\cite{WHH}
The value of $\mu_{0}H_{c2}$(0) exceeds the Pauli paramagnetic
limit\cite{Hp} [$H_{P}$(0)$\approx$1.84$T_c$ tesla for an isotropic
s-wave spin-singlet superconductor] by 35\%. Similar observations
have been reported in LaFeAsO$_{1-x}$F$_{x}$\cite{Fuchs} and
LaFe$_{1-x}$Ni$_{x}$AsO\cite{Ni} systems. The upper critical field
of LaFePO is far below its Pauli paramagnetic limit,\cite{Hamlin}
revealing the difference between LaFePO and LaFeAs$_{0.7}$P$_{0.3}$O
superconductors.

\begin{figure}
\onefigure[width=8cm]{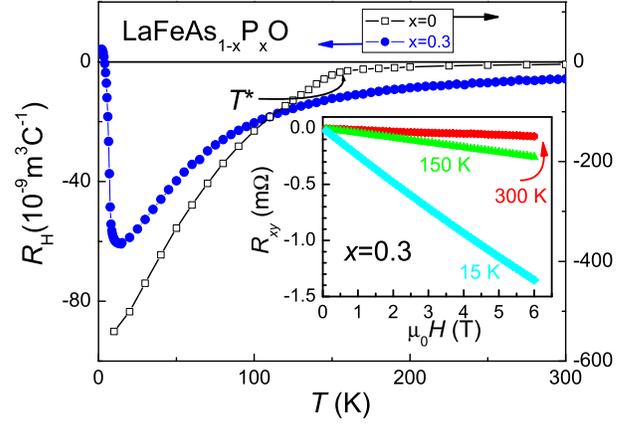} \caption{Temperature dependence of
Hall coefficient for the superconductor LaFeAs$_{0.7}$P$_{0.3}$O, in
comparison with that of the parent compound LaFeAsO. The inset shows
the field dependence of the Hall resistance at several fixed
temperatures.}\label{fig.6}
\end{figure}

Fig.~\ref{fig.6} shows that the Hall coefficient ($R_H$) of
LaFeAs$_{0.7}$P$_{0.3}$O is negative in the normal state, suggesting
the dominant charge transport by the electron conduction. The normal
state $R_H(T)$ exhibits very strong temperature dependence
(especially at low temperatures), compared with the
LaFeAsO$_{1-x}$F$_{x}$ superconductors.\cite{Oak,ZhuZW} This
indicates that the multiband effect is more significant in the
P-doped superconductors. The room temperature $R_H$ value is
$-$5.7$\times$10$^{-9}$m$^3$C$^{-1}$, very close to that of the
undoped LaFeAsO ($-$4.8$\times$10$^{-9}$m$^3$C$^{-1}$), supporting
that the P doping does not induce extra charge carriers. Below 10 K,
$|R_H|$ decreases very sharply, consistent with the superconducting
transition. The non-zero $|R_H|$ is due to the non-zero resistance
under high magnetic fields. The $|R_H|$ of undoped LaFeAsO increases
rapidly below $T^* \sim$155 K because of the structural phase
transition and the subsequent SDW ordering.\cite{Oak} Such a
transition cannot be detected in LaFeAs$_{0.7}$P$_{0.3}$O. Since the
P doping does not change the number of Fe 3$d$ electrons, the severe
suppression of SDW order by P doping suggests that Fermi surface
nesting is unlikely to account for the SDW ordering in the LaFeAsO.

Now, let's discuss the occurrence of superconductivity in P-doped
LaFeAsO. While the chemical-pressure-induced superconductivity in
P-doped LaFeAsO basically agrees with the static-pressure-induced
superconductivity in LaFeAsO,\cite{P-La1111} there is difference
between the two kinds of pressure. The hydrostatic pressure
generally produces a homogeneous effect, but chemical pressure may
be selective to a particular structural unit in a complex compound.
In the present LaFeAs$_{1-x}$P$_{x}$O system, P-doping leads to the
squeezing (stretching) in Fe$_{2}$As$_{2}$ (La$_{2}$O$_{2}$) layers,
respectively, along the $c$-axis. Band calculations\cite{Band}
suggest that the relative positions of As/P to Fe planes affect the
electronic structure. When arsenic is moved closer to the iron
planes, the two-dimensional pocket with $d_{xy}$ character in the
LaFeAsO evolves into a three-dimensional pocket with
$d_{3z^{2}-r^{2}}$ character. This might correlate with the
three-dimensional superconductivity in the two-dimensional
system.\cite{Yuan} Thus, the appearance of superconductivity in
P-doped LaFeAsO suggests that $d_{3z^{2}-r^{2}}$ band should be
important for superconductivity. It is noted that the La-site
replacement with smaller rare earth elements, which also produce
chemical pressures, influences little on the SDW order, and induces
no superconductivty.\cite{Ln1111} A possible reason is that the
chemical pressure is applied mainly in La$_{2}$O$_{2}$ rather than
Fe$_{2}$As$_{2}$ layers.

In addition to the above structural points of view, the difference
in covalency for the bonding of Fe-As/P may also play a role for
superconductivity. The P substitution for As in the undoped iron
pnictides was proposed as a means to access the magnetic quantum
criticality in an unmasked fashion.\cite{JDai} The narrow
superconducting region in LaFeAs$_{1-x}$P$_{x}$O supports the
scenario of quantum criticality with magnetic fluctuations for the
superconducting mechanism. Nevertheless, much work is needed to
address this issue.

In summary, we have discovered bulk superconductivity in
LaFeAs$_{1-x}$P$_{x}$O by the isovalent substitution of As with P.
Superconductivity emerges in the region of $0.2<x<0.4$ with the
maximum $T_c$ of 10.8 K. Unlike previous doping strategy in LaFeAsO,
the P doping does not change the number of Fe 3$d$ electrons. This
chemical-pressure-induced superconductivity in ferroarsenides
contrasts sharply with the case in cuprates, where superconductivity
is always induced by doping of charge carriers into an AFM Mott
insulator. We suggest that both the chemical pressure (applied
selectively to the Fe$_{2}$As$_{2}$ layers) and the covalency of
Fe-P bonding may lead to quantum criticality, facilitating
superconductivity.

\acknowledgments{This work is supported by the NSF of China
(Contract Nos. 10674119 and 10634030), National Basic Research
Program of China (Contract No. 2007CB925001) and the PCSIRT of the
Ministry of Education of China (IRT0754).}

\end{document}